\documentclass[12pt]{article}
\usepackage[utf8]{inputenc}
\usepackage[pdftex]{graphicx}
\usepackage{hyperref}
\usepackage{amsfonts,amssymb,amsmath}

\newtheorem{defin}{Definition}
\newtheorem{lemma}{Lemma}
\newtheorem{theorem}{Theorem}
\newtheorem{remark}{Remark}
\newtheorem{proposition}{Proposition}

\newcommand{\HESS}{{\bf Hess}}
\newcommand{\K}{{\bf K}}
\newcommand{\Hess}{{\rm Hess}}
\newcommand{\J}{J}
\newcommand{\Tr}{{\rm Tr}}

\title{On the detailed structure of quantum control landscape for fast single qubit phase-shift gate generation}
\author{Boris O. Volkov$^{1,2}\footnote{\url{www.mathnet.ru/eng/person/94935}}$ and Alexander N. Pechen$^{1,2}$\footnote{\url{www.mathnet.ru/eng/person/17991}}}

\begin{document}

\maketitle

$^1$ Department of Mathematical Methods for Quantum Technologies, Steklov Mathematical Institute of Russian Academy of Sciences, 8 Gubkina Str., Moscow, 119991, Russia\footnote{\url{www.mi-ras.ru/eng/dep51}}\\
$^2$ Quantum Engineering Research and Education Center, University of Science and Technology MISIS, 4 Leninsky Prosp., Moscow 119991, Russia\\

E-mail: borisvolkov1986@gmail.com, apechen@gmail.com (corresponding author)

\begin{abstract}
    In this work, we study the detailed structure of quantum control landscape for the problem of single-qubit phase shift gate generation on the fast time scale. In previous works, the absence of traps for this problem was proved on various time scales. A special critical point which was known to exist in quantum control landscapes was shown to be either a saddle or a global extremum, depending on the parameters of the control system. However, in the case of saddle the numbers of negative and positive eigenvalues of Hessian at this point and their magnitudes have not been studied. At the same time, these numbers and magnitudes determine the relative ease or difficulty for practical optimization in a vicinity of the critical point. In this work, we compute the numbers of negative and positive eigenvalues of Hessian at this saddle point and moreover, give estimates on magnitude of these eigenvalues. We also significantly simplify our previous proof of the theorem about this saddle point of the Hessian [Theorem~3 in B.O.~Volkov, O.V.~Morzhin, A.N.~Pechen, J.~Phys.~A: Math. Theor. {\bf 54}, 215303 (2021)].
\end{abstract}

\section{Introduction}  
Optimal quantum control, which includes methods for manipulation of quantum systems, attracts now high attention due to various existing and prospective applications in quantum technologies~\cite{GlaserEurPhysJD2015,KochEPJ1,ButkovskyBook1990, TannorBook2007, LetokhovBook2007,MooreCS2011,KochJPhysCondensMatter2016,AlessandroBook2021}. Among important topics, one problem which was posed in~\cite{RHR} is the analysis of quantum control landscapes, that is, local and global extrema of quantum control objective functionals. Various results have been obtained in this field e.g. in~\cite{HR,MoorePRA2011,Pechen2011,Pechen2012,IJC2012,FouquieresSchirmer,PechenIl'in2014,PechenTannorCJC2014,Larocca2018,Zhdanov2018,Russell2018,PechenIl'in2016,Volkov_Morzhin_Pechen,DalgaardPRA2022} for closed and open quantum systems. For open quantum systems, a formulation of completely positive trace preserving dynamics as points of {\it complex Stiefel manifold} (strictly speaking, of some factors of complex Stiefel manifolds over some equivalence relation) was proposed and theory of open system's quantum control as gradient flow optimization over complex Stiefel manifolds was developed in details for two-level~\cite{PechenJPA2008} and general $n$--level quantum systems~\cite{OzaJPA2009} and applied to the analysis of quantum control landscapes. Control landscapes for open-loop and closed-loop control were analyzed in a unified framework~\cite{PechenPRA2010.82.030101}. A unified analysis of classical and quantum kinematic control landscapes was performed~\cite{PechenEPJ2010}. Computation of numbers of positive and negative eigenvalues of the Hessian at saddles of the control landscape is an important problem~\cite{MoorePRA2011}.

For numerical optimization in quantum control, various local and global search methods are used including such as based on Pontryagin maximum principle~\cite{PontryaginBook1962,BoscainPRXQuantum2021}, GRadient Ascent Pulse Engineering (GRAPE)~\cite{khaneja_optimal_2005}, gradient flows~\cite{Glaser2010}, Krotov type methods~\cite{Tannor1992,Morzhin2019}, gradient free CRAB optimisation~\cite{CanevaPRA2011}, Hessian based methods such as the Broyden–Fletcher–Goldfarb–Shanno (BFGS) algorithm~\cite{EitanPRA2011}, geometric methods~\cite{AgrachevBook2004}, genetic algorithms~\cite{Judson1992},  machine learning~\cite{DongIEEE2008}, dual annealing~\cite{MorzhinLJM2021}, etc. The importance of mathematical analysis of extrema of quantum control objective functionals is motivated by the fact that local but not global maxima (if they would exist) would impede the search for globally optimal controls using local search algorithms, which could be more efficient otherwise. For this reason, points of local but not global extrema of the objective functional are called {\it traps}.

In this work, we analytically study the detailed structure of quantum control landscape around special critical point for the problem of single-qubit phase shift gate generation on the fast time scale. The absence of traps for single-qubit gate generation was proven on long time scale in~\cite{Pechen2012,PechenIl'in2014} and on fast time scale~\cite{PechenIl'in2016,Volkov_Morzhin_Pechen}, where a single special control which is a critical point was studied and shown to be a saddle. However, the numbers of negative and positive eigenvalues of Hessian at this saddle point control were not  studied. At the same time, these numbers are important as they determine the numbers of directions towards decreasing and increasing of the objective and hence determine the level of difficulty for practical optimization starting in a vicinity of the saddle point. The numbers of positive and negative eigenvalues of Hessian of the objective for some other examples of quantum systems were computed in~\cite{MoorePRA2011}.  In this work, we compute the numbers of negative and positive eigenvalues of Hessian at this saddle point, give estimates on magnitude of the eigenvalues and also significantly simplify our previous proof of the theorem about Hessian at this saddle point~\cite{Volkov_Morzhin_Pechen}. Numerical experiments for  the problem of single-qubit phase shift gate generation on the fast time scale were performed in~\cite{Volkov_Morzhin_Pechen,Morzhin_Pechen_arxive}.

In Sec.~\ref{Sec:2}, we summarize results of previous works which are relevant for our study. In Sec.~\ref{Sec:3}, the main theorem of this work is presented; its proof is provided in Sec.~\ref{Sec:4}. Conclusions Sec.~\ref{Sec:5} summarizes this work.

\section{Previous results}\label{Sec:2}
A single qubit driven by a coherent control $f\in L^2([0,T];\mathbb R)$, where $T>0$ is the final time, in the absence of the environment evolves according to the Schr\"odinger equation for the unitary evolution operator $U_t^f$
\begin{equation}
\frac{dU_t^f}{dt}=-i(H_0+f(t)V)U_t^f,\qquad U_0^f=\mathbb I
\end{equation}
where $H_0$ and $V$ are the free and interaction Hamiltonians. Common assumption is that $[H_0,V]\ne 0$. This assumption guarantees controllability of the two-level system for large time; otherwise the dynamics is trivial. In this work, we consider time scale smaller than the controllability time. A single qubit quantum gate is a unitary $2\times 2$ operator $W$ defined up to a physically unrelevant phase, so that $W\in SU(2)$. The problem of single qubit gate generation can be formulated as
\begin{equation}\label{eq:JW}
\J_W[f]=\frac14 |\Tr (U_T^f W^\dagger)|^2\to\max
\end{equation}

\begin{defin} 
Control $f^*$ is called trap for the problem~(\ref{eq:JW}) if $f^*$ is a point of local but not global maximum of $J_W$, i.e. $J(f^*)<\sup\limits_f J_W[f]$.
\end{defin}

In~\cite{PechenIl'in2014,PechenIl'in2016,Volkov_Morzhin_Pechen},  results on the absence of traps for this problem were obtained. To explicitly formulate these results, consider the special constant control $f(t)=f_0$ and time $T_0$:
\begin{eqnarray}
f_0&:=&\frac{-\Tr H_0\Tr V+2\Tr(H_0V)}{(\Tr V^2)^2-2\Tr(V^2)},\\
T_0&:=&\frac {\pi}{\|H_0-\mathbb I \Tr H_0/2+f_0(V-\mathbb I  \Tr V/2)\|},
\end{eqnarray}
where $\|\cdot\|$ denotes the spectral norm.

The following theorem was proved in~\cite{PechenIl'in2014}. 

\begin{theorem}\label{theorem2016-1}
Let $W\in SU(2)$ be a single qubit quantum gate.
If $[W,H_0+f_0V]\neq0$ then for any $T>0$ traps do not exist. If $[W,H_0+f_0V]=0$ then any control, except possibly $f\equiv f_0$, is not trap for any $T>0$ and the control  $f_0$ is not  trap for $T>T_0$.
\end{theorem}

The case of whether control $f_0$ can be trap for $T\leq T_0$ or not was partially studied in~\cite{PechenIl'in2016}. Without loss of generality it is sufficient to consider the case $H_0=\sigma_z$ and $V=v_x\sigma_x+v_y\sigma_y$, where $\upsilon_x,\upsilon_y\in \mathbb{R}$ ($\upsilon_x^2+\upsilon_y^2>0$) and $\sigma_x,\sigma_y,\sigma_z$ are the Pauli matrices:
\begin{equation}
\sigma_x=\Bigg(\begin{array}{*{20}{c}}0 & 1 \\ 1 & 0\end{array}\Bigg), \qquad  \sigma_y=\Bigg(\begin{array}{*{20}{c}}0 & -i \\ i & 0\end{array}\Bigg), \qquad \sigma_z=\Bigg(\begin{array}{*{20}{c}} 1 & 0 \\ 0 & -1\end{array}\Bigg).
\end{equation}
In this case, the special time is $T_0=\frac {\pi}2$ and the special control is $f_0=0$.

By Theorem~\ref{theorem2016-1}, if $[W,\sigma_z]\neq 0$, then  for any  $T>0$ there are no traps for $J_W$. If $[W,\sigma_z]=0$, then $W=e^{i\varphi_W \sigma_z+i\beta}$, where  $\varphi_W\in (0,\pi]$ and $\beta\in [0,2\pi)$. The phase can be neglected, so without loss of generality we set $\beta=0$. Below we consider only such gates. The following result was proved in~\cite{PechenIl'in2016}. 

\begin{theorem}\label{theorem2016-2}
Let $W=e^{i\varphi_W \sigma_z}$. If $\varphi_W\in (0,\frac{\pi}{2})$, then for any $T>0$ there are no traps. If $\varphi_W\in [\frac \pi 2,\pi]$, then for any $T>\pi-\varphi_W$ there are no traps.
\end{theorem}

For fixed $\varphi_W$ and $T$ the value of the objective evaluated at $f_0$ is
\begin{equation}\label{JD2}
J_W[f_0]=\cos^2{(\varphi_W+T)}.
\end{equation}
If $\varphi_W+T=\pi$ then $J_W[f_0]=1$ and $f_0$ is a point of global maximum. If $\varphi_W+T=\frac{\pi}2$ and $\varphi_W+T=\frac{3\pi}2$ then $J_W[f_0]=0$ and $f_0$ is a point of  global minimum. 

The Taylor expansion of the functional $J_W$ at $f$ up to the second order has the form (for the theory of calculus of variations in infinite dimensional spaces see~\cite{BogachevSmolyanov}):
\begin{equation}
 J_W[f+\delta f] = J_W[f]+J_W^{(1)}[f](\delta f)+\frac 12J_W^{(2)}[f](\delta f,\delta f) +o(\|\delta f\|^{2})\text{ as $\|\delta f\|\rightarrow 0$.}
\end{equation}
The first Fr\'echet derivative is $$J_W^{(1)}[f](\delta f)=\int_0^T\frac{\delta J_W}{\delta f(t)}\delta f(t)dt,$$
 where  the integral kernel, which determines the gradient of the objective, is 
\[
\frac{\delta J_W}{\delta f(t)}=\frac 12 \Im(\Tr Y^\ast \Tr(YV_t))
\]
Here as in~\cite{PechenIl'in2016} we use the notations $Y=W^\dagger U^f_T$ and $V_t=U^{f\dagger}_t VU^f_t$. The second order term is
\begin{equation}
\frac 12J_W^{(2)}[f](\delta f,\delta f)=\frac12(\HESS\, \delta f, \delta f)_{L^2}
=\frac12\int_0^T\int_0^T\Hess (t,s)f(t)f(s)dtds, 
\end{equation}
where Hessian $\HESS \colon L^2([0,T],\mathbb{R})\to L^2([0,T],\mathbb{R})$ is an  integral operator:
\begin{equation}
(\HESS f)(t)=\int_0^T\Hess (t,s)f(s)ds.
\end{equation}
The integral kernel of the Hessian has the form
\[
\Hess(t,s)=
\begin{cases}
\frac 12 \operatorname{Re}(\Tr(YV_{t})\Tr(Y^\ast V_{s})-\Tr(YV_{s}V_{t})\Tr Y^\ast)
,&\textrm{ if } s\geq t\\
\frac 12 \operatorname{Re}(\Tr(YV_{s})\Tr(Y^\ast V_{t})-\Tr(YV_{t}V_{s})\Tr Y^\ast)
,& \textrm{ if } s<t.
\end{cases}
\]
The control $f_0=0$ is a critical point, i.e., gradient of the objective evaluated at this control is zero. The integral kernel of Hessian at $f_0=0$ has the form (see~\cite{PechenIl'in2016}):
\begin{equation}\label{Hess}
\Hess(s,t)=-2\upsilon^2\cos{\varphi}\cos{(2|t-s|+\varphi)},
\end{equation}
where $\varphi=-\varphi_W-T$ and $\upsilon=\sqrt{\upsilon^2_x+\upsilon^2_y}$.

We consider for the values of the parameters $(\varphi_W,T)$ the following cases (see Fig.~\ref{fig2}, where the set $\mathcal{D}_2$ in addition is divided into three subsets described in Sec.~\ref{Sec:4}):
\begin{itemize}
\item $(\varphi_W,T)$ belongs to the triangle domain
\[
\hspace*{-1.2cm}
\mathcal{D}_1 := \left\{ (\varphi_W, T)~:~ 0<T<\frac{\pi}{2}, \quad \frac{\pi}{2} \leq \varphi_W < \pi - T \right\}; 
\] 
\item $(\varphi_W,T)$ belongs to the triangle domain
\[
\hspace*{-1.2cm}
\mathcal{D}_2 := \left\{ (\varphi_W, T)~:~ 0 < T \leq \frac{\pi}{2}, \quad \pi - T < \varphi_W < \pi, \quad  (\varphi_W,T)\neq (\pi, \frac {\pi}2) \right\}.
\]
\item $(\varphi_W,T)$ belongs to the square domain without the diagonal
\[
\hspace*{-1.2cm}
\mathcal{D}_3 := \left\{(\varphi_W, T)~:~ 0 < T \leq \frac{\pi}{2}, \quad 0 < \varphi_W <\frac {\pi}2 ,\quad  \varphi_W+T\neq\frac{\pi}2\right\}.
\]
\item $(\varphi_W,T)$ belongs to the set
\[
\hspace*{-1.2cm}
\mathcal{D}_4 := \left\{ (\varphi_W, T)~:~ 0 < T \leq \frac{\pi}{2},\quad \varphi_W=\pi \right\}.
\]
\end{itemize}
\begin{remark}
Note that these notations for the domains $\mathcal{D}_2,\mathcal{D}_3,\mathcal{D}_4$ are different from that used in~\cite{Volkov_Morzhin_Pechen}. The present notations seem to be more convenient. 
\end{remark}

The following theorem was obtained in~\cite{Volkov_Morzhin_Pechen}.
\begin{theorem}
If $(\varphi_W,T)\in \mathcal{D}_1\cup \mathcal{D}_2\cup \mathcal{D}_3\cup \mathcal{D}_4$ then the Hessian of the objective functional $J_W$ at $f_0=0$ is an injective compact  operator on $L^2([0,T];\mathbb{R})$. Moreover, the following holds.
\begin{enumerate}
\item If $(\varphi_W,T)\in \mathcal{D}_1$, then Hessian at $f_0$ has only 
negative eigenvalues.
\item If $(\varphi_W,T)\in \mathcal{D}_2\cup \mathcal{D}_3\cup \mathcal{D}_4$  then Hessian 
at $f_0$ has both negative and positive eigenvalues. In this case, the special control $f_0=0$ is a saddle point for the objective functional. 
\end{enumerate}
\end{theorem}

Numerical experiments suggest that if $(\varphi_W,T)\in{\cal D}_1$ then $f_0$ is a point of global maximum~\cite{Morzhin_Pechen_arxive}. Note that the numbers of positive and negative eigenvalues mentioned in item 2 of this theorem, as well as their magnitudes, were not computed in~\cite{Volkov_Morzhin_Pechen}. However, these numbers and magnitudes are important since they determine the numbers of directions towards increasing or decreasing of $J_W$, and the magnitudes of the eigenvalues determine the speed of increasing and decreasing the objective along these directions. All of that affects the relative level of easy or difficulty of practical optimization in a vicinity of $f_0$.

\section{Main theorem}\label{Sec:3}
Our main result of this work is the following theorem. 
\begin{theorem}\label{theorem4} One has the following.
\begin{enumerate}
\item If $(\varphi_W,T)\in \mathcal{D}_1$, then Hessian at $f_0$ has only 
negative eigenvalues.
\item If $(\varphi_W,T)\in \mathcal{D}_2$,  then Hessian 
at $f_0$ has two positive and infinitely many negative eigenvalues.  
\item  If $(\varphi_W,T)\in\mathcal{D}_3$ and $\varphi_W+T<\pi/2$, 
 then Hessian at $f_0$ has one positive and infinitely many negative eigenvalues.
If $(\varphi_W,T)\in\mathcal{D}_3$ and $\varphi_W+T>\pi/2$, then Hessian at $f_0$ has one negative and infinitely many positive eigenvalues.
\item If $(\varphi_W,T)\in\mathcal{D}_4$,  then Hessian at $f_0$ has one negative and infinitely many positive eigenvalues.
\end{enumerate}
\end{theorem}

\section{Proof of the main theorem}\label{Sec:4}

In this section, we will investigate the spectrum of Hessian $\HESS$ and prove Theorem~4.

If $(\varphi_W,T)\in \mathcal{D}_1\cup \mathcal{D}_2\cup \mathcal{D}_3\cup \mathcal{D}_4$, then $\sin2\varphi=-\sin2(\varphi_W+T)\neq 0$.
Instead of Hessian, we can consider the operator ${\bf K}=\frac 1{\upsilon^2\sin2\varphi} \HESS$ which differs from $\HESS$ by a scalar factor that makes the calculations a bit simpler. Here ${\bf K}\colon L^2([0,T],\mathbb{R}) \to L^2([0,T],\mathbb{R})$ is an integral operator:
$$
(\K f)(t)=\int_0^TK(t,s)f(s)ds
$$
with the integral kernel
\begin{equation}
K(t,s)=-\frac{\cos{(2|t-s|+\varphi)}}{\sin{\varphi}},
\end{equation}
Because operators $\K$  and $\HESS$ differ by a scalar factor,  their spectra are related
$$
\sigma(\K)=\frac 1{\upsilon^2\sin2\varphi}\sigma(\HESS).
$$
Let $g\in L^2([0,T],\mathbb{R})$ and $h=\K g$. Then
\begin{align}
\label{importeq}
  h(t)=-\frac 1{\sin{\varphi}}\int_0^t\cos{(2t-2s+\varphi)}g(s)ds-\frac 1{\sin{\varphi}}\int_t^T\cos{(2s-2t+\varphi)}g(s)ds.  
\end{align}
Differentiating twice in a generalized sense  expression~(\ref{importeq}), we obtain (see~\cite{Volkov_Morzhin_Pechen} for details)
\begin{equation}
\label{eq1!}
h''(t)+4h(t)=4g(t).
\end{equation}
Moreover, for  any continuous  $g$, we can find  $h=\K g$ as a unique solution of ODE~(\ref{eq1!}),
 which satisfies the initial conditions
\begin{equation}
\label{bound1}
h(0)=-\frac 1{\sin{\varphi}}\int_0^T\cos{(2s+\varphi)}g(s)ds,
\end{equation}
\begin{equation}
\label{bound2}
h'(0)=-\frac {2}{\sin{\varphi}}\int_0^T\sin{(2s+\varphi)}g(s)ds. 
\end{equation}
These initial conditions are obtained by substituting $t=0$ into the right hand side of expression~(\ref{importeq})  and into its derivative.

Equality~(\ref{eq1!}) implies that if $h\equiv 0$ then  $g\equiv 0$. Hence $\K$ is an injective operator.
Let $\mu\neq 0$ be an eigenvalue of the operator $\K$ and $g$ be a corresponding eigenfunction, then $h=\K g=\mu g$. 
Let $\lambda=1/\mu$.
Then~(\ref{eq1!}),~(\ref{bound1}) and~(\ref{bound2}) together imply that
the search for eigenvalues of the operator $\K$ reduces to the problem of finding such 
$h\in C^\infty([0,T],\mathbb{R})$ and nonzero $\lambda \in\mathbb{R}$ that
\begin{equation}\label{hh}
\begin{cases}
h''(t)=4(\lambda-1)h(t),\\
\lambda h(0)=-\frac 1{\sin{\varphi}}\int_0^T\cos{(2s+\varphi)}h(s)ds,\\
\lambda h'(0)=-\frac {2}{\sin{\varphi}}\int_0^T\sin{(2s+\varphi)}h(s)ds.
\end{cases}
\end{equation}
This problem is similar to the Sturm–Liouville  problem (see~\cite{Vladimirov}).

\subsection{Case $\lambda<1$}
Consider the case of nonzero $\lambda<1$.
Let $a^2=(1-\lambda)$ and $a>0$. If $h$ satisfies~(\ref{hh}) then $h$ has the form $$h(t)=b\cos 2at+c\sin 2at.$$
If we substitute $h$ in the boundary conditions of~(\ref{hh}), then we get a system of two linear algebraic equations on $(b,c)$. It has a non-zero solution if the determinant of the coefficients of this system is not equal to zero. This determinant has the form (see~\cite{Volkov_Morzhin_Pechen}) 
\[
 F^1_{\varphi_W,T}(a)=-2a-a^2\sin{(2aT)}\sin{(2\varphi_W)}-\sin{(2aT)}\sin{(2\varphi_W)}+2a\cos{(2aT)}\cos({2\varphi_W)}.
\]
So the function $h$  and $\lambda<1$ are a solution of problem~(\ref{hh}) if and only if
the function $F^1_{\varphi_W,T}$ has a positive root.

It is easy to see that for $(\varphi_W,T)\in\mathcal{D}_4$
the roots of the function $F^1_{\varphi_W,T}$ are  $a_n=\frac{\pi n}T$. If $(\varphi_W,T)\in\mathcal{D}_1$
and $\varphi_W=\frac{\pi}2$ then the roots of the function $F^1_{\varphi_W,T}$ are  $a_n=\frac{(2n-1)\pi}{2T}$. Hence, in these cases,  $\mu_n=\frac 1{1-a_n^2}$ belong to the spectrum of the operator $\K$.  We will show bellow that there is also only one positive eigenvalue  for $(\varphi_W,T)\in\mathcal{D}_4$ and there are not negative eigenvalues for $(\varphi_W,T)\in\mathcal{D}_1$
such that $\varphi_W=\frac{\pi}2$.

\begin{figure}[ht]\label{fig1}
\includegraphics[width=1\linewidth]{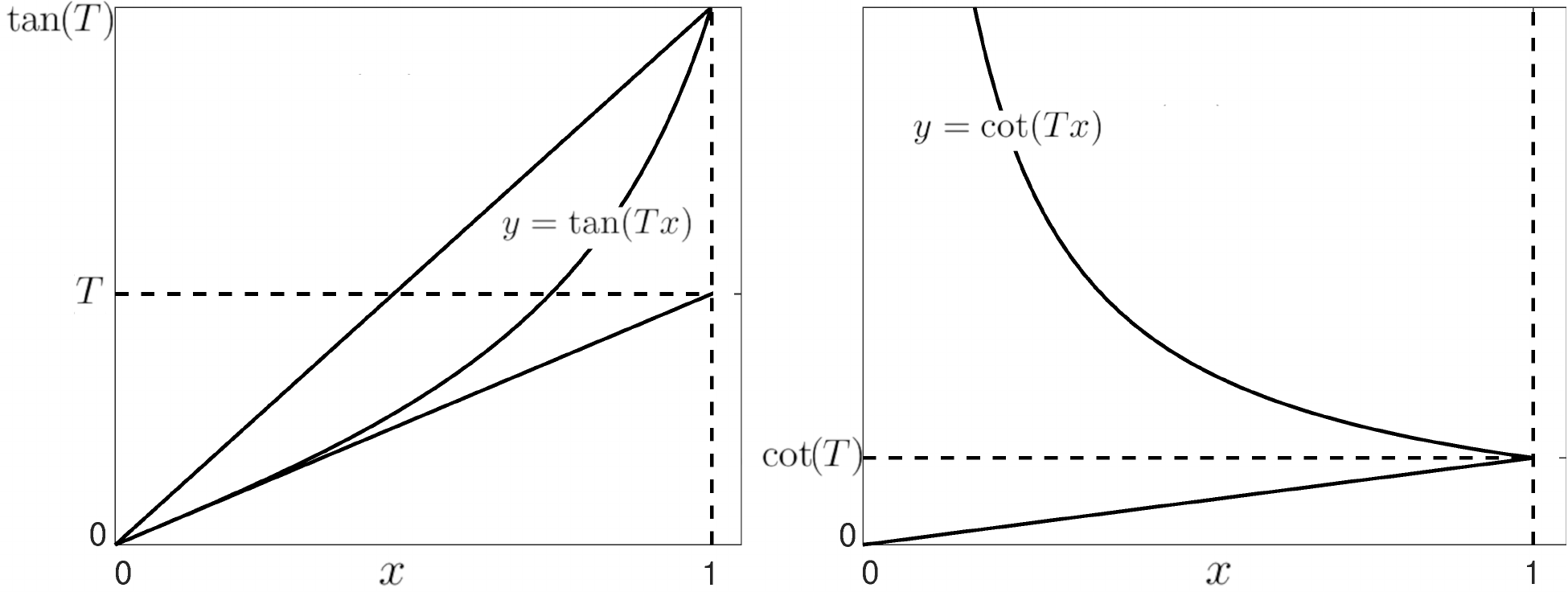}
\caption{Illustration for Lemma~\ref{lemma2} (left) and for Lemma~\ref{lemma3} (right). Left: Two inclined straight lines have angles of inclination $T$ and $\tan(T)$, i.e., they are defined by equations $y=Tx$ and $y=\tan(T)x$, respectively. The line $y=\tan(Tx)$ on the interval $(0,1)$ lies between these two strait lines. Hence on the interval  $(0,1)$ line $y=\alpha x$ intersects line $y=\tan(Tx)$ if and only if $T<\alpha<\tan(T)$.
Right: The inclined straight line $y=\cot(T)x$ intersects with line $y=\cot(Tx)$ at $x=1$. Hence equation $\alpha x=\cot(Tx)$ on $(0,1)$ has one solution if $\alpha> \cot{T}$ and has no solutions if $\alpha\le\cot{T}$.}
\end{figure}

\begin{lemma}
\label{lemma2}
Let $T\in(0,\frac\pi 2)$. The equation
\[
\alpha x=\tan{(Tx)}
\]
has only one root on $(0,1)$ if $T< \alpha <\tan{T}$ and has not roots on $(0,1)$ if $\alpha\in(-\infty,T]$ and $\alpha \in[\tan{T},+\infty)$.
\end{lemma}
The proof of this lemma is illustrated on Fig.~\ref{fig1} (left subplot).

\begin{lemma}
\label{lemma3}
Let $T\in(0,\frac\pi 2)$. The equation
\[
\alpha x=\cot{(Tx)}
\]
has only one root on $(0,1)$ if $\cot{T}< \alpha$ and has not roots on $(0,1)$ if $\alpha\in(-\infty,\cot{T}]$.
\end{lemma}
The proof of this lemma is illustrated on Fig.~\ref{fig1} (right subplot).

\begin{figure}[ht]\label{fig2}
\centering
\includegraphics[width=.75\linewidth]{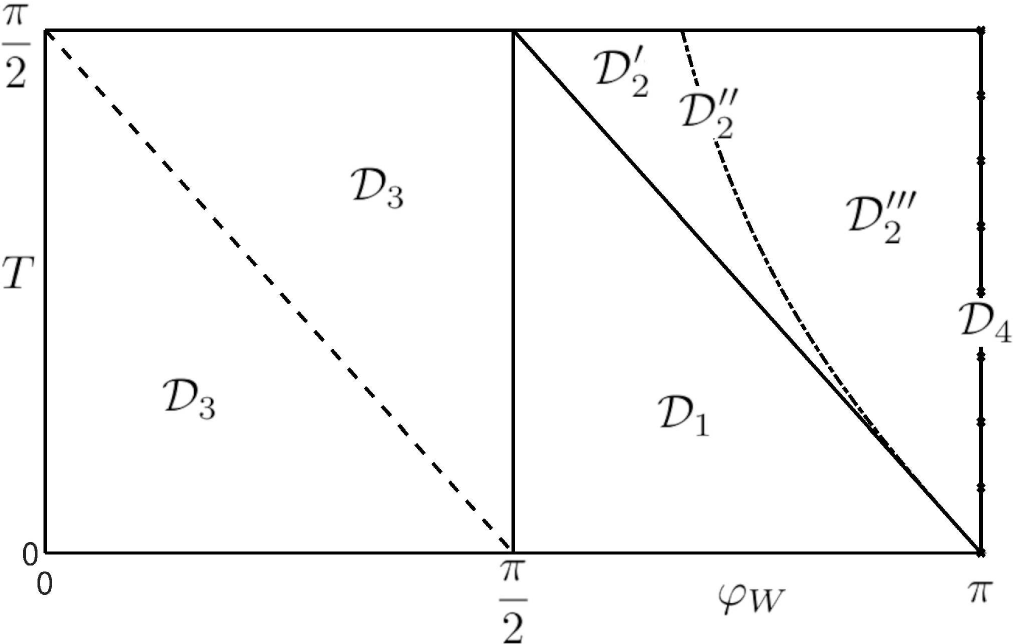}
\caption{The domains of the rectangle $(\varphi_W,T)\in[0,\pi]\times [0,\pi/2]$. $\mathcal{D}_3$ is the left square except of the dashed diagonal and borders. $\mathcal{D}_1$ is the bottom triangle in the right square with left vertical border and without the solid diagonal and bottom horizontal border. $\mathcal{D}_4$ is the vertical line marked with crosses. $\mathcal{D}''_2$ is the dash-dotted line. $\mathcal{D}_2'$ is the area between $\mathcal{D}_1$ and $\mathcal{D}_2''$. $\mathcal{D}_2'''$ is the area between $\mathcal{D}_2''$ and $\mathcal{D}_4$. Note that these notations for the domains $\mathcal{D}_2,\mathcal{D}_3,\mathcal{D}_4$ are different from that used in~\cite{Volkov_Morzhin_Pechen}.
In the domain $\mathcal{D}_1$, the Hessian at critical point $f_0=0$ has only 
negative eigenvalues. On the dashed diagonal (in the left square), the critical point $f_0=0$ is a point of global minimum. On the solid diagonal (in the right square), the critical point $f_0=0$ is a point of global maximum. On the domain $\mathcal{D}_2=\mathcal{D}_2'\cup \mathcal{D}_2''\cup \mathcal{D}_2'''$, the critical point $f_0=0$   is a saddle point with  two positive and infinitely many negative eigenvalues of Hessian. On the bottom triangle of $\mathcal{D}_3$,  the critical point $f_0=0$   is a saddle point with one positive and infinitely many negatives eigenvalues of the Hessian. On the top triangle of $\mathcal{D}_3$ and on the domain $\mathcal{D}_4$,  the critical point $f_0=0$ is a saddle point with one negative and infinitely many positive eigenvalues of the Hessian.
}
\end{figure}

In addition, we divide the domain ${\cal D}_2$ into the following three subdomains (see Fig.~\ref{fig2}).
\begin{itemize}
\item  $(\varphi_W,T)$ belongs to the set
\[
\hspace*{-1.2cm}
\mathcal{D}'_2 := \left\{ (\varphi_W, T)~:~(\varphi_W,T)\in \mathcal{D}_2,\quad  T<-\tan{\varphi_W} \right\};
\]
\item  $(\varphi_W,T)$ belongs to the set
\[
\hspace*{-1.2cm}
\mathcal{D}_2'':= \left\{ (\varphi_W, T)~:~(\varphi_W,T)\in \mathcal{D}_2,\quad T=-\tan{\varphi_W} \right\};
\]
\item  $(\varphi_W,T)$ belongs to the set
\[
\hspace*{-1.2cm}
\mathcal{D}_2''' := \left\{ (\varphi_W, T)~:~(\varphi_W,T)\in \mathcal{D}_2,\quad T>-\tan{\varphi_W} \right\}.
\]
\end{itemize}

\begin{proposition}\label{prop1}
Positive roots of the function $F^1_{\varphi_W,T}$ are
positive solutions $\{a'_m\}$ and $\{a_n\}$  of the equation
\begin{equation}
\label{eq2}
-x\tan{\varphi_W}=\tan{(xT)}
\end{equation}
and the equation
\begin{equation}
\label{eq1}
-x\cot{\varphi_W}=\cot{(xT)}
\end{equation}
respectively. Here $\{a'_m\}$ and $\{a_n\}$ are two countable sets  whose elements are distributed in ascending order.   Then  the numbers $\mu'_m=\frac 1{1-{a'}^2_m}$  and $\mu_n=\frac 1{1-a_n^2}$,  belong to the spectrum of the operator ${\bf K}$. Moreover, the following holds.

\begin{enumerate}
\item If $(\varphi_W,T)\in \mathcal{D}_1$ and $\varphi_W\neq \frac{\pi}2$,  then
 $a'_m\in \left(\frac{(m-1)\pi}{T},\frac{(2m-1)\pi}{2T}\right)$ and $a_n\in \left(\frac{(n-1)\pi}{T},\frac{(2n-1)\pi}{2T}\right)$, where $n,m\in \mathbb{N}$. In this case, the numbers $\{\mu'_m\}$ and $\{\mu_n\}$  are negative for all
$n,m\in\mathbb{N}$.
\item If $(\varphi_W,T)\in \mathcal{D}''_2\cup \mathcal{D}'''_2$,  then $a'_m\in \left(\frac{(m-1)\pi}{T},\frac{(2m-1)\pi}{2T}\right)$  and $a_n\in \left(\frac{(n-1)\pi}{T},\frac{(2n-1)\pi}{2T}\right)$,  where $m\in \{2,3,\ldots\}$ and  $n\in \mathbb{N}$.   In this case, $\mu_1=\frac 1{1-a_1^2}>1$ is positive and the numbers  $\{\mu'_m\}$ and $\{\mu_n\}$ are negative for $n>1$ and $m>1$. 
\item If $(\varphi_W,T)\in  \mathcal{D}'_2$, then
 $a'_m\in \left(\frac{(m-1)\pi}{T},\frac{(2m-1)\pi}{2T}\right)$ and $a_n\in \left(\frac{(n-1)\pi}{T},\frac{(2n-1)\pi}{2T}\right)$, where $n,m\in \mathbb{N}$. In this case,   $\mu'_1=\frac 1{1-{a'_1}^2}>1$ and $\mu_1=\frac 1{1-a_1^2}>1$  are  positive.  The numbers  $\{\mu'_m\}$ and $\{\mu_n\}$ are negative for $n>1$ and $m>1$.
\item If $(\varphi_W,T)\in \mathcal{D}_3$, then  
  $a'_m\in \left(\frac{(2m-1)\pi}{2T},\frac{m\pi}{T}\right)$ and $a_n\in \left(\frac{(2n-1)\pi}{2T},\frac{n\pi}{T}\right)$, where  $n,m\in \mathbb{N}$. In this case, the numbers   $\{\mu'_m\}$ and $\{\mu_n\}$ are negative for all $n,m\in\mathbb{N}$.
\end{enumerate}
\end{proposition}
\textbf{Proof}
Let us analyze positive roots of the function $F^1_{\varphi_W,T}$.
For this purpose we consider quadratic (with respect to $x$) equation:
\[
x^2\sin{(2aT)}\sin{(2\varphi_W)}+2x(1-\cos{(2aT)}\cos{(2\varphi_W)})+\sin{(2aT)}\sin{(2\varphi_W)}=0
\]
The roots of this quadratic equation are
\begin{eqnarray}
x_1&=&-\cot{\varphi_W}\tan{\left(aT\right)},\\
x_2&=&-\tan{\varphi_W}\cot{\left(aT\right)}
\end{eqnarray}
Hence  $a$ is a root of the function $F^1_{\varphi_W,T}$ 
if and only if $x=a$ is a solution of either  equation ~(\ref{eq2}) or equation~(\ref{eq1}). 
If $(\varphi_W,T)\in \mathcal{D}_3$, then  both equations  have a single root on each interval
  $\left(\frac{(2n-1)\pi}{2T},\frac{n\pi}{T}\right)$ for $n\in\mathbb{N}$. If $(\varphi_W,T)\in \mathcal{D}_1\cup \mathcal{D}_2$ and $\varphi_W\neq \frac{\pi}2$, then  equation~(\ref{eq1})  has a single root on each interval
  $\left(\frac{(n-1)\pi}{T},\frac{(2n-1)\pi}{2T}\right)$ for $n\in\mathbb{N}$.  If $(\varphi_W,T)\in \mathcal{D}_1\cup\mathcal{D}'_2$ and $\varphi_W\neq \frac{\pi}2$, then  equation~(\ref{eq2}) has a single root on each interval
  $\left(\frac{(n-1)\pi}{T},\frac{(2n-1)\pi}{2T}\right)$ for $n\in\mathbb{N}$. Lemma~\ref{lemma2}  implies that, in the case $(\varphi_W,T)\in \mathcal{D}''_2\cup\mathcal{D}'''_2$,  equation~(\ref{eq2})  has not roots on  $\left(0,\frac{\pi}{2T}\right)$ and has  a single root on each interval
  $\left(\frac{(n-1)\pi}{T},\frac{(2n-1)\pi}{2T}\right)$ for $n>1$.

 The number $\mu=\frac 1{1-a^2}$
is a positive eigenvalue of the operator $\bf K$ if and only if
$a\in (0,1)$. If $(\varphi_W,T)\in \mathcal{D}_1$ and $\varphi_W\neq \frac{\pi}2$ then 
\[
\tan{T}<\tan{(\pi-\varphi_W)}=-\tan{(\varphi_W)}.
\]
Due to Lemma~\ref{lemma3} equation~(\ref{eq1}) has not roots on   $(0,1)$.
Due to Lemma~\ref{lemma2} equation~(\ref{eq2})
has not roots on $(0,1)$.

If $(\varphi_W,T)\in \mathcal{D}_2$, then
\[
\cot{T}<-\cot{\varphi_W}.
\]
Hence, due to Lemma~\ref{lemma3} equation~(\ref{eq1}) has one root on   $(0,1)$.
If $T\geq -\tan{\varphi_W}$ then  Lemma~\ref{lemma2}   
implies that~(\ref{eq2}) has not solution on $(0,1)$.
So if $(\varphi_W,T)\in \mathcal{D}''_2\cup \mathcal{D}'''_2$, then the function $F^1_{\varphi_W,T}$ has only one root $a_1$ on the interval $(0,1)$.
If $(\varphi_W,T)\in \mathcal{D}'_2$, then the function $F^1_{\varphi_W,T}$ has two roots $a_1$ and $a'_1$ on the interval $(0,1)$.

If $(\varphi_W,T)\in \mathcal{D}_3$, then $-\tan{\varphi_W}<0$ and Lemmas~\ref{lemma2} and~\ref{lemma3} imply that  both equations~(\ref{eq2}) and~(\ref{eq1})  have not solutions on $(0,\frac{\pi}{2T})$. Hence, the numbers   $\{\mu'_m\}$ and $\{\mu_n\}$ are negative for all $n,m\in\mathbb{N}$.

\subsection{Case $\lambda=1$}
\label{4.2}
If $\mu=1$ is an eigenvalue of the operator ${\bf K}$ then
the corresponding eigenfunctions should have the form
\[
h(t)=g(t)=ct+b.
\]

If we substitute $g$ in~(\ref{bound1}) and~(\ref{bound2}), then we get a system of two linear algebraic equations on $(b,c)$. This system has a non-zero solution if and only if the determinant of the coefficients of this system is not equal to zero. 
This determinant has the form (see~\cite{Volkov_Morzhin_Pechen})
\begin{equation}
\Delta=-2\sin{\varphi_W}(\sin{\varphi_W}+T\cos{\varphi_W}).
\end{equation}

\begin{proposition}\label{prop2}
$\mu'_1=1$ is an eigenvalue
of ${\bf K}$ only in the following cases
\begin{enumerate}
\item  $(\varphi_W,T)\in \mathcal{D}_{4}$.
\item  $(\varphi_W,T)\in \mathcal{D}_2''$.
\end{enumerate}
\end{proposition}

\subsection{Case $\lambda>1$}

Consider the case $\lambda>1$.
Let $a^2=(\lambda-1)$ and $a>0$. If $h$ satisfies~(\ref{hh}) then $h$ has the form
\[
h(t)=be^{2at}+ce^{-2at}.
\]
If we substitute $h$ in boundary conditions of~(\ref{hh}), then we get a system of two linear algebraic equations on $b$ and $c$. This system has a non-zero solution if and only if the determinant of the coefficients of this system is not equal to zero. 
This determinant has the form~\cite{Volkov_Morzhin_Pechen} 
\begin{eqnarray}
\label{1111bound11} 
F^2_{\varphi_W,T}(a) &=&-a^2\sinh{(2aT)}\sin{(2\varphi_W)} \nonumber \\
&&+2a(1-\cosh{(2aT)}\cos{(2\varphi_W)})
+\sinh{(2aT)}\sin{(2\varphi_W)}.
\end{eqnarray}
So the function $h$  and $\lambda>1$ are a solution of problem~(\ref{hh}) if and only if
the function $F^2_{\varphi_W,T}$ has a positive root.

It is easy to see that for $(\varphi_W,T)\in\mathcal{D}_4$
and $(\varphi_W,T)\in\mathcal{D}_1$ such that $\varphi_W=\frac{\pi}2$  the function $F^2_{\varphi_W,T}$ has not positive roots.

\begin{proposition}\label{prop3} One has the following.
\begin{enumerate}
\item If $(\varphi_W,T)\in \mathcal{D}_3$, then $F^2_{\varphi_W,T}$ has only one positive root $a'_0>0$, where $a'_0$ is a solution of the equation
\begin{equation}
\label{eq12}
x\cot{\varphi_W}=\coth{Tx}
\end{equation}
Then $\mu'_0=\frac1{1+a_0^2}<1$ is a positive eigenvalue of the operator ${\bf K}$.
\item  
If $(\varphi_W,T)\in \mathcal{D}_1\cup \mathcal{D}'_2\cup \mathcal{D}''_2$ and $\varphi_W\neq \frac{\pi}2$, then the function  $F^2_{\varphi_W,T}$ has no positive roots.
\item If $(\varphi_W,T)\in \mathcal{D}'''_2$,
 then $F^2_{\varphi_W,T}$ has only one positive root $a'_1>0$, where $a'_1$ is a solution of the equation
 \begin{equation}
\label{eq11}
-x\tan{\varphi_W}=\tanh{Tx}.
\end{equation}
 Then $\mu'_1=\frac1{1+{a'_1}^2}<1$
is a positive eigenvalue of the operator ${\bf K}$.
\end{enumerate}
\end{proposition}
\textbf{Proof.}
Let us analyze positive roots of the function $F^2_{\varphi_W,T}$.
For this purpose we consider quadratic (with respect to $x$) equation:
\[
x^2\sinh{(aT)}\sin{(2\varphi_W)}-2x(1-\cosh{(aT)}\cos{(2\varphi_W)})-\sinh{(aT)}\sin{(2\varphi_W)}=0
\]
The roots of this equation are 
\begin{eqnarray}
x_1&=&-\cot{\varphi_W}\tanh{\left(aT\right)}\\
x_2&=&\tan{\varphi_W}\coth{\left(aT\right)}
\end{eqnarray}

Then $a$ is a root of the function $F^2_{\varphi_W,T}$ 
if and only if $x=a$ is a solution of either equation~(\ref{eq12})
or equation~(\ref{eq11}).
 If $(\varphi_W,T)\in \mathcal{D}_3$ then due to $\tan{\varphi_W}>0$ equation~(\ref{eq11}) has not positive roots. Equation~(\ref{eq12}) has only one positive root.

 If $(\varphi_W,T)\in \mathcal{D}_1\cup \mathcal{D}_2$ and $\varphi_W\neq \frac \pi 2$ then due to $\cot{\varphi_W}<0$ equation~(\ref{eq12}) has not positive roots. If  $(\varphi_W,T)\in \mathcal{D}_1\cup \mathcal{D}'_2\cup \mathcal{D}''_2$ and $\varphi_W\neq \frac \pi 2$
then $T\leq-\tan\varphi_W$ and $\tanh{(Tx)}<-\tan({\varphi_W})x$ for positive $x$ and equation~(\ref{eq11}) has not positive roots.

 If $(\varphi_W,T)\in \mathcal{D}'''_2$ then  equation~(\ref{eq11}) has one positive root.

The statement of Theorem~\ref{theorem4} follows directly from Propositions~\ref{prop1},~\ref{prop2},~\ref{prop3}. Important is that these propositions in addition give estimates for the magnitudes of the eigenvalues.

\section{Conclusions} 
\label{Sec:5}
Analysis of either existence or absence of traps (which are points of local, but not global, extrema of the objective quantum functional) is important for quantum control. It was known that in the problem of single-qubit 
phase shift quantum gate  generation all controls, except maybe the special control $f_0=0$ at small times, cannot be traps. In the previous work~\cite{Volkov_Morzhin_Pechen}, we studied the spectrum of the Hessian at this control $f_0$ and investigated under what conditions this control is a saddle point of the quantum objective functional. In this work, we have calculated the numbers of negative and positive eigenvalues of the Hessian at this control point and obtained estimates for the magnitudes of these eigenvalues. At the same time, we significantly simplified the proof of Theorem~3 of the paper~\cite{Volkov_Morzhin_Pechen}.

\section*{Acknowledgements} 
Authors thank A.\,I.~Mikhailov for helpful discussions and advice. This work was funded by Russian Federation represented by the Ministry of Science and Higher Education (grant number 075-15-2020-788).

\bibliographystyle{unsrt}
\bibliography{VolkovPechenLandscape2.bib}

\begin{thebibliography}{10}

\bibitem{GlaserEurPhysJD2015}
S.J. Glaser, U.~Boscain, T.~Calarco, C.P. Koch, W.~K\"{o}ckenberger, R.~Kosloff, I.~Kuprov, B.~Luy, S.~Schirmer, T.~Schulte-Herbr\"{u}ggen, D.~Sugny, and F.K. Wilhelm.
\newblock Training {S}chr\"{o}dinger's cat: quantum optimal control. strategic report on current status, visions and goals for research in europe.
\newblock {\em The European Physical Journal~D}, 69(12), 2015.
\newblock Article no.~279.

\bibitem{KochEPJ1}
C.P. Koch, U.~Boscain, T.~Calarco, G.~Dirr, S.~Filipp, S.J. Glaser, R.~Kosloff, S.~Montangero, T.~Schulte-Herbr\"{u}ggen, D.~Sugny, and F.K. Wilhelm.
\newblock Quantum optimal control in quantum technologies. strategic report on current status, visions and goals for research in europe.
\newblock {\em EPJ Quantum Technol.}, 9, 2022.
\newblock Article no.~19.

\bibitem{ButkovskyBook1990}
A.~G. Butkovskiy and Y.~I. Samoilenko.
\newblock {\em Control of Quantum-Mechanical Processes and Systems}.
\newblock Nauka Publ., Moscow (in russian), eng. transl.: Kluwer Acad. Publ., Dordrecht, 1984, 1990.

\bibitem{TannorBook2007}
D.~J. Tannor.
\newblock {\em Introduction to Quantum Mechanics: A Time Dependent Perspective}.
\newblock Univ. Science Books, Sausilito, CA, 2007.

\bibitem{LetokhovBook2007}
V.~Letokhov.
\newblock {\em Laser Control of Atoms and Molecules}.
\newblock Oxford Univ. Press, 2007.

\bibitem{MooreCS2011}
K.~W. Moore, A.~Pechen, X.~J. Feng, J.~Dominy, V.~J. Beltrani, and H.~Rabitz.
\newblock Why is chemical synthesis and property optimization easier than expected?
\newblock {\em Physical Chemistry Chemical Physics}, 13(21):10048--10070, 2011.

\bibitem{KochJPhysCondensMatter2016}
C.~P. Koch.
\newblock Controlling open quantum systems: Tools, achievements, and limitations.
\newblock {\em J. Phys.: Condens. Matter}, 28(21):213001, 2016.

\bibitem{AlessandroBook2021}
D.~D'Alessandro.
\newblock {\em Introduction to Quantum Control and Dynamics}.
\newblock Chapman \& Hall, Boca Raton, 2nd edition, 2021.

\bibitem{RHR}
H.A. Rabitz, M.M. Hsieh, and C.M. Rosenthal.
\newblock Quantum optimally controlled transition landscapes.
\newblock {\em Science}, 303(5666):1998--2001, 2004.

\bibitem{HR}
T.-S. Ho and H.~Rabitz.
\newblock Why do effective quantum controls appear easy to find?
\newblock {\em Journal of Photochemistry and Photobiology~A: Chemistry}, 180(3):226--240, 2006.

\bibitem{MoorePRA2011}
K.W. Moore, R.~Chakrabarti, G.~Riviello, and H.~Rabitz.
\newblock Search complexity and resource scaling for the quantum optimal control of unitary transformations.
\newblock {\em Phys. Rev. A}, 83:012326, 2011.

\bibitem{Pechen2011}
A.N. Pechen and D.J. Tannor.
\newblock Are there traps in quantum control landscapes?
\newblock {\em Phys. Rev. Lett.}, 106(12), 2011.
\newblock Article no.~120402.

\bibitem{Pechen2012}
A.~Pechen and N.~Il'in.
\newblock Trap-free manipulation in the {L}andau-{Z}ener system.
\newblock {\em Phys. Rev.~A}, 86(5), 2012.
\newblock Article no.~052117.

\bibitem{IJC2012}
A.N. Pechen and D.J. Tannor.
\newblock Quantum control landscape for a lambda-atom in the vicinity of second-order traps.
\newblock {\em Israel Journal of Chemistry}, 52(5):467--472, 2012.

\bibitem{FouquieresSchirmer}
P.~De~Fouquieres and S.G. Schirmer.
\newblock A closer look at quantum control landscapes and their implication for control optimization.
\newblock {\em Infinite Dimensional Analysis, Quantum Probability and Related Topics}, 16(3), 2013.
\newblock Article no.~1350021.

\bibitem{PechenIl'in2014}
A.N. Pechen and N.B. Il'in.
\newblock Coherent control of a qubit is trap-free.
\newblock {\em Proceedings of the Steklov Institute of Mathematics}, 285(1):233--240, 2014.

\bibitem{PechenTannorCJC2014}
A.N. Pechen and D.J. Tannor.
\newblock Control of quantum transmission is trap-free.
\newblock {\em Canadian Journal of Chemistry}, 92(2):157--159, 2014.

\bibitem{Larocca2018}
M.~Larocca, P.M. Poggi, and D.A. Wisniacki.
\newblock Quantum control landscape for a two-level system near the quantum speed limit.
\newblock {\em J. Phys.~A: Math. Theor.}, 51(38), 2018.
\newblock Article no.~385305.

\bibitem{Zhdanov2018}
D.V. Zhdanov.
\newblock Comment on '{C}ontrol landscapes are almost always trap free: a geometric assessment'.
\newblock {\em J. Phys. A: Math. Theor.}, 51, 2018.
\newblock Article no.~508001.

\bibitem{Russell2018}
B.~Russell, R.~Wu, and H.~Rabitz.
\newblock Reply to comment on '{C}ontrol landscapes are almost always trap free: a geometric assessment'.
\newblock {\em J. Phys. A: Math. Theor.}, 51, 2018.
\newblock Article no.~508002.

\bibitem{PechenIl'in2016}
A.N. Pechen and N.B. Il'in.
\newblock On extrema of the objective functional for short-time generation of single-qubit quantum gates.
\newblock {\em Izv. Math.}, 80(6):1200--1212, 2016.

\bibitem{Volkov_Morzhin_Pechen}
B.~Volkov, O.~Morzhin, and A.~Pechen.
\newblock Quantum control landscape for ultrafast generation of single-qubit phase shift quantum gates.
\newblock {\em J. Phys. A: Math. Theor.}, 54, 2021.
\newblock Article no.~215303.

\bibitem{DalgaardPRA2022}
M.~Dalgaard, F.~Motzoi, and J.~Sherson.
\newblock Predicting quantum dynamical cost landscapes with deep learning.
\newblock {\em Phys. Rev. A}, 105:012402, 2022.

\bibitem{PechenJPA2008}
A.~Pechen, D.~Prokhorenko, R.~Wu, and H.~Rabitz.
\newblock Control landscapes for two-level open quantum systems.
\newblock {\em J. Phys. A.: Math. Theor.}, 41(4):045205, 2008.

\bibitem{OzaJPA2009}
A.~Oza, A.~Pechen, J.~Dominy, V.~Beltrani, K.~Moore, and H.~Rabitz.
\newblock Optimization search effort over the control landscapes for open quantum systems with {K}raus-map evolution.
\newblock {\em J. Phys. A: Math. Theor.}, 42(20):205305, 2009.

\bibitem{PechenPRA2010.82.030101}
A.~Pechen, C.~Brif, R.~Wu, R.~Chakrabarti, and H.~Rabitz.
\newblock General unifying features of controlled quantum phenomena.
\newblock {\em Phys. Rev. A}, 82:030101, 2010.

\bibitem{PechenEPJ2010}
A.~Pechen and H.~Rabitz.
\newblock Unified analysis of terminal-time control in classical and quantum systems.
\newblock {\em Europhysics Letters}, 91(6):60005, 2010.

\bibitem{PontryaginBook1962}
L.S. Pontryagin, V.G. Boltyanskii, R.V. Gamkrelidze, and E.F. Mishchenko.
\newblock {\em The Mathematical Theory of Optimal Processes / Transl. from Russian}.
\newblock Interscience Publishers John Wiley \& Sons, Inc., New York--London, 1962.

\bibitem{BoscainPRXQuantum2021}
U.~Boscain, M.~Sigalotti, and D.~Sugny.
\newblock Introduction to the pontryagin maximum principle for quantum optimal control.
\newblock {\em PRX Quantum}, 2, 2021.
\newblock Article no.~030203.

\bibitem{khaneja_optimal_2005}
N~Khaneja, T~Reiss, C~Kehlet, T~Schulte-Herbrüggen, and S.J. Glaser.
\newblock Optimal control of coupled spin dynamics: design of {NMR} pulse sequences by gradient ascent algorithms.
\newblock {\em J. Magn. Reson.}, 172(2):296--305, 2005.

\bibitem{Glaser2010}
T.~Schulte-Herbr\"uggen, S.~J. Glaser, G.~Dirr, and U.~Helmke.
\newblock Gradient flows for optimization in quantum information and quantum dynamics: foundations and applications.
\newblock {\em Rev. Math. Phys.}, 22:597--667, 2010.

\bibitem{Tannor1992}
D.J. Tannor, V.~Kazakov, and V.~Orlov.
\newblock Control of photochemical branching: novel procedures for finding optimal pulses and global upper bounds.
\newblock {\em Time-dependent quantum molecular dynamics, Nato ASI Ser. Ser. B}, 299:347--360, 1992.

\bibitem{Morzhin2019}
O.~V. Morzhin and A.~N. Pechen.
\newblock Krotov method for optimal control of closed quantum systems.
\newblock {\em Russian Mathematical Surveys}, 74(5):851--908, oct 2019.

\bibitem{CanevaPRA2011}
T.~Caneva, T.~Calarco, and S.~Montangero.
\newblock Chopped random-basis quantum optimization.
\newblock {\em Phys. Rev. A}, 84:022326, 2011.

\bibitem{EitanPRA2011}
R.~Eitan, M.~Mundt, and D.~J. Tannor.
\newblock Optimal control with accelerated convergence: Combining the krotov and quasi-newton methods.
\newblock {\em Phys. Rev. A}, 83:053426, 2011.

\bibitem{AgrachevBook2004}
A.A. Agrachev and Y.L. Sachkov.
\newblock {\em Control Theory from the Geometric Viewpoint}.
\newblock Springer, 2004.

\bibitem{Judson1992}
R.~S. Judson and H.~Rabitz.
\newblock Teaching lasers to control molecules.
\newblock {\em Phys. Rev. Lett.}, 68:1500--1503, 1992.

\bibitem{DongIEEE2008}
Daoyi Dong, Chunlin Chen, Tzyh-Jong Tarn, A.~Pechen, and H.~Rabitz.
\newblock Incoherent control of quantum systems with wavefunction-controllable subspaces via quantum reinforcement learning.
\newblock {\em IEEE Transactions on Systems, Man, and Cybernetics, Part B (Cybernetics)}, 38(4):957--962, 2008.

\bibitem{MorzhinLJM2021}
O.V. Morzhin and A.N. Pechen.
\newblock Generation of density matrices for two qubits using coherent and incoherent controls.
\newblock {\em Lobachevskii J. Math.}, 42(10):2401--2412, 2021.

\bibitem{Morzhin_Pechen_arxive}
O.V. Morzhin and A.N. Pechen.
\newblock On optimization of coherent and incoherent controls for two-level quantum systems.
\newblock {\em Izv. Math.}, 87(5):1024--1050, 2023.

\bibitem{BogachevSmolyanov}
V.I. Bogachev and O.G. Smolyanov.
\newblock {\em Topological Vector Spaces and Their Applications. Springer Monographs in Mathematics}.
\newblock Springer, 2017.

\bibitem{Vladimirov}
V.S. Vladimirov.
\newblock {\em Equations of Mathematical Physics}.
\newblock Mir Publisher, Moscow, 2nd edition, 1983.

\end{thebibliography}

\end{document}